\documentclass[12pt]{article}
\usepackage[english]{babel}
\usepackage{a4wide}
\usepackage{latexsym}
\date{}

\begin{document}

\title{\bf {\large{Arguments for F-Theory}}}
\author{\normalsize{Luis J. Boya \footnote{luisjo@unizar.es}} \\
\normalsize{Departamento de F\'{\i}sica Te\'{o}rica} \\
\normalsize{Universidad de Zaragoza} \\
\normalsize{E-50009 Zaragoza} \\
\normalsize{SPAIN}} 

\maketitle

\begin{abstract}

After a brief review of string and $M$-Theory we point out some
deficiencies. Partly to cure them, we present several arguments
for ``$F$-Theory'', enlarging spacetime to $(2, 10)$ signature,
following the original suggestion of C. Vafa. We introduce a
suggestive Supersymmetric  $27$-plet of particles, associated to
the exceptional symmetric hermitian space $E_{6}/Spin^{c}(10)$.
Several possible future directions, including using projective
rather than metric geometry, are mentioned. We should emphasize
that $F$-Theory is yet just a very provisional attempt, lacking
clear dynamical principles. \\



\end{abstract}

\vfill \eject

\section {Introduction to M theory.}

\subsection {The beginning of String Theory.}

Strings entered high-energy physics ``from the back door'' around
1970, as a theoretical explanation for the Veneziano formula found
to fit hadron resonances. In the first period of string theory, up
to 1984/5, several advances were made: fermions were included,
with supersymmetry first in the worldsheet (Gervais and Sakita),
then, after the Gliozzi-Scherk-Olive (GSO) projection (1977), also
in spacetime (in $10 = (1, 9)$ dimensions); string interactions
were studied through vertex operators; etc. \\

Open strings include massless spin one particles (``$A$'' fields)
as simple excitations, conducive of gauge theories, and closed
strings (including the closed sector of open strings) exhibit also
a massless spin two excitation (``$h$'') which resembled the
graviton; besides, a two-form (``$B$''), radiated by the string,
is always present, together with a scalar (``$\phi$''), the
dilaton, all still in the massless sector. The scale of the
massive excitations is set up by the string constant (or Regge slope)
$\alpha^{\prime}$, or $T = 1/(2 \pi \alpha^{\prime}$ ) with dim $
T = (length)^{-2}$. By 1975 it was clear that hadron physics was
better explained by QCD (Quantum ChromoDynamics), a Yang-Mills
theory with $SU(3)$ color as gauge group; also about the same time
Scherk and Schwarz \cite{Schs} proposed to apply string
theory to gravitation and other forces because, for example,
the previous $s=2$ and $s=1$ excitations had interactions, as
expected, reminiscent of gravitation and gauge couplings; the
proposal caught the physics community totally surprised and
unconvinced. \\

String theory enjoyed very smooth high-energy behaviour (in
particular gravitational interactions seemed to be tamed for the
first time); string theory was also found to be free of anomalies 
(abundant in higher dimension spaces), or even better, the gauge
groups and the spacetime dimensions were selected precisely by the
absence of anomalies. Even the worrysome tachyons of the primitive
bosonic string (living in 26 dimensions) disappeared in the GSO
supersymmetric version (living in ten). \\

On the other hand, the Standard Model (SM) was completed also
around 1975, and included just the three gauge couplings for
electromagnetic, weak and strong forces. Moreover,
these couplings run at different speed, to nearly coincide around
$10^{15}$ GeV \cite{Geor}, to reinforce the Grand Unified Theory (GUT);
and this was not very far from the Planck scale,
$10^{19}$ GeV, where supposedly gravitational forces also became
important. So it was reassuring that both string theory (through
the extant gravitons) and the SM (through running couplings) led
to schemes which foresaw gravitation close to other forces, for
the first time realistically in the history of physics. For an
early review of string theory see \cite{Schr}, and also
\cite{Gso} for target supersymmetry. \\

\subsection {The Five Superstring Theories.}

By 1985 it was clear that there were just five definite consistent
superstring theories, with the following characteristics: all of them
 live in ten ($=1,9$) dimensions, all were supersymmetric both
on the worldsheet and in the target space, the five included
gravitons, and some include gauge fields, the ones with  $N=1$
target Supersymmetry, see below. \\

There is one open string with gauge group $O(32)$ (called Type $I$),
two closed $N=2$ superstrings, called $IIA$ and $IIB$, and
another two closed $N=1$ ``heterotic'' strings with gauge groups
$O(32)$ (Het-Ortho or H-O) and $E_8 \times E_8$ (called Het-Excep
or H-E), where $E_8$ is the compact form of the last exceptional
group of Cartan, with $248$ generators. \\

The following scheme reminds the situation \\

\begin{equation}\label{eq:0}
\begin{array}{clcr}

IIA &       & Het-Excep, H-E \\
    &       &            \\
IIB &        &     Het-Ortho, H-O  \\
    &       &    \\
    &   Type I   &
\end{array}
\end{equation}
\\

A short description follows. We start by Type $I$; it is a theory
of open (and closed) strings, with $N = 1$ SuperSymmetry, and with a gauge
group $O(32)$. The fundamental supersymmetry is the dimension $8$
($8$ transverse directions) equivalence between the vector $\Box $
and the (chiral) spinor $\Delta_{L,R}$ representations of
$O(8)(dim \ \ spinor =8
= 2^{8/2-1}$, type +1, real). \\

\begin{equation}\label{eq:00}
\begin{array}{clcr}
\Box  &   -    & \Delta \\
8 &    -    &     8
\end{array}
\end{equation}

The particle content of this Type I theory in the $m = 0$ limit
includes the fundamental gauge particle (gaugeon, $32 \cdot
31/2=496$ of them, plus gaugino(s)), and the closed sector content is
obtained from the \textit{skew} square of (2), to wit

\begin{equation}\label{eq:000}
\begin{array}{clcr}
(\Box - \Delta) \wedge  (\Box - \Delta) &   =    & h + \phi + B - ( \Psi + \psi) \\
(8 - 8) \wedge^2 &    =    &     35 + 1 + 28 - (56 +8)
\end{array}
\end{equation}

\noindent (where $\Psi$ is the gravitino field and $\psi$
represents a spinor field). Notice the three essential
ingredients, namely graviton, dilaton and two-form in the Bose
sector, plus gravitino and spinors in the Fermi sector. The
dilaton $\phi$ is crucial for the perturbative expansions, because
in string theory $\exp(\phi)$ plays the role of the coupling
constant $g = g_s$: perturbation theory is topological, and one
expands in the number
of ``holes'' in the worldsheet as a bidimensional surface. \\

In the Heterotic - Orthogonal corner, which is a closed string, the
particle content is the same as the closed $+$ open sector of the
previous case. On the other hand, the Heterotic-Exceptional corner
has the product  group $E_8 \times E_8$ \textit{en lieu}  of the
$O(32)$ orthogonal group; notice both groups have the same
dimension (496) and rank (16). \\

$N=2$  Supersymmetry in target space implies closed strings only,
with no gauge groups. The difference lies in the two chiralities,
which can be equal ($IIB$ theory, chiral) or parity transformed
($IIA$ theory, nonchiral). The massless particle content is just
the square of the fundamental equality (2)

\begin{equation}\label{eq:0000}
\begin{array}{clcr}
IIA: (\Box - \Delta_{L}) \times  (\Box - \Delta_{R}) &   =    &
 \ \ \ \ \ h +  B + \phi + A + C  - (2 \Psi + 2 \psi) \\
   &   =  &  35 + 28 + 1 + 8 + 56 - 128
\end{array}
\end{equation}

\begin{equation}\label{eq:00000}
\begin{array}{clcr}
IIB: (\Box - \Delta_{L, R}) \times  (\Box - \Delta_{L, R}) &   =    &  \
\ \ \ \ h +  B + \phi + \phi' + B' + D^{\pm}
- (2 \Psi + 2 \psi) \\
   &   =  &  35 + 28 + 1 + 1 + 28 + 35 - 128
\end{array}
\end{equation}

\noindent (where $D^{\pm}$ is the (anti-)selfdual 4-form). The
bose particles in the $NS$ ( Neveu-Schwarz) sector come from the
bose $\times$ bose product, whereas the $RR$ (Ramond) sector
is the fermi $\times$ fermi. This distinction is very important. \\

It was also early noticed that the particle content of the maximal
supergravity (SuGra, Sugra), namely $11D$ Sugra \cite{Crems} gives the
$m=0$ content of the IIA string after dimensional reduction:

\begin{equation}\label{eq:a}
\begin{array}{clcr}
11D \ \  SuGra:  &   h - \Psi + C \\
    &    44 - 128 + 84
\end{array}
\end{equation}

\noindent with the $(D + 1 ) \rightarrow D$ decomposition rules
for the graviton $h$ and 3-form $C$

\begin{equation}\label{eq:b}
h_{D+1} = h_{D} + A_{D} + \phi_{D}, \ \ \ \ C_{D+1} = C_{D} +
B_{D}
\end{equation}

\noindent and for the gravitino $\Psi $

\begin{equation}\label{eq:c}
\begin{array}{clcr}
\Psi_{11}  &  = &  \Psi_{10,L} + \Psi_{10,R} + \psi_{10,R} + \psi_{10,L} \\
128     &  = & 56 \ \  + 56 \ \ + 8 \ \ +8
\end{array}
\end{equation}

The GSW book \cite{Gsw} is the best reference for this period.

\subsection { The first  ``Theory Of  Everything, T.O.E.''}

How far were we at the time from real nature? For one thing, extra
dimensions ($6=10-4$) supposed not to be seen in $4D$ because the six
were compact with radius of order of elementary (= Planck\'{}s)
length; for another, SuperSymmetry had to be badly broken, as the
naive,``toroidal'' compactification would lead to unwanted $N=8$ or $N=4$
Supersymmetry in $4D$.  In particular, compactification and Susy
breaking could be together. \\

Thus the Heterotic Exceptional corner of the above pentagon
(1) enjoyed a temporary splendor, around 1985, as the best candidate
for a realistic theory \cite{Cand}: compactification in a
Calabi-Yau 3-fold (i.e., a complex three dimensional K\"{a}hler and
Ricci-flat manifold or orbifold) with Euler number $\chi = \pm 6$ could
be `close' to generate a model of particles including even the
three generations. One has to insist on $N=1$ only Susy in
$4D$, as to allow parity violating couplings. It was to be
hoped to get gauge forces from the descent chain of groups and subgroups

\begin{equation}\label{eq:1}
E_{8} \times E_{8} \rightarrow  E_{8} \ ( + \  hidden \ \  sector)
\ \rightarrow E_{6}
\end{equation}

\noindent as $E_6$ is the biggest of the three best candidates for
a Grand Unified Theory (GUT): these groups are $E_6$, $SO(10)=E_5$ and
$SU(5)=E_4$: they allow for complex
representations and are big enough to be simple but still
encompass the groups of the Standard Model. \\

These early succeses are well told in the book \cite{Gsw}; the
term ``Theory Of Everything'' aroused around that time \cite{Dav},
with some opposition from more ``conventional'' physicists, who
strongly opposed this quasi-theological approach! \cite{Ginsg}.
\\

From the phenomenological point of view, it is also to be remarked
that improved renormalization group calculations, including
minimal suspersymmetry, reproduce better the common intersection of the three
coupling constants at somewhat higher energy, around $10^{16}$ GeV \cite{Dine}. \\

\subsection { M Theory.}

In 1995, ten years later, three confluent developments brought
about the so-called $M$-theory: it turned out that all five SuperString
theories were related, and also with the mentioned maximal $11D$
supergravity; there were also higher dimension extended objects
($p$-Branes, $p > 1$, with $ p = 0$ for particles, $p = 1$ for
strings, etc.), and several nonperturbative results were obtained;
besides, as the theory had no adjustable parameters, it seemed to
be potentially selfsufficient. The hopes for a unique theory were
higher than ever; many previously skeptical physicists (including
this humble reviewer) were converted to the superstring \textit{Credo}. \\

These developments have to to with dualities (E. Witten
\cite{W95}), membranes (P. Townsend \cite{To94}) and $D$-$p$-branes
(J. Polchinski \cite{Pol95}). \\

The equivalences take place mainly through the concepts of
$T$-duality and $S$-duality. $T$-duality (\cite{Give}) is
characteristic of strings, where the \textit{winding} modes, with
energy proportional to the radius, exchange with \textit{momentum}
modes, with energy inverse with radius: selfduality, with enhanced
symmetry, occurs at the special radius $r = \alpha^{´\prime}/r$.
$T$-duality relates $IIA$ and $IIB$ theories, as the rolling
interchanges chirality; it also related the two heterotic strings;
recall the two groups $E_{8}^{2}$ and $O(32)$ share the same
dimension and rank. The status of $T$-duality is rather robust, as
it is satisfied order by order in perturbation theory.  \\

$S$-duality is more conjectural but more deep; it was first stated
explicitely around 1990 \cite{Font}, and generalized by Witten and
others. It related one theory at a coupling $g$ with another theory
at coupling $1/g$; it was inspired in some models in quantum field
theory, mainly Coleman's proof of the equivalence of the
sine-Gordon theory to the massive Thirring model with reciprocal
couplings, and the Montonen-Olive conjectures about
electromagnetic charge-monopole duality.  Also the strong limit of
strings shows the existence of ``solitonic" objects, like
membranes, and in general $D-p$-branes (Polchinski), which are
endstations for open strings. In particular for a longtime a
membrane was seen floating in eleven dimensional gravity, so P.
Townsend in particular hypothetized, at some early date, that the
$11D$ Sugra \textit{cum} membrane theory was just a better, eleven
dimensional version of the ($IIA$) string theory. The scheme was
this: \\

\begin{equation}\label{eq:19}
\begin{array}{clcr}
11D \ \ Membranes &  \rightarrow  & 11D \ Sugra \\
	&						&        \\
?? &   & dim  \  \downarrow \ reduc  \\
	&				&				\\
IIA \ \ Theory &   \ \ \rightarrow & IIa \ Sugra  \\
  &    m = 0  &
\end{array}
\end{equation}

In other important prescient finding, Strominger showed in 1990
\cite{Stro90} that the $7 = 10 - 3$ -dimensional dual field
strength extant in $10D$ strings was really coupled to a solitonic
$5$-Brane, dual to the fundamental string $1$-Brane. Some of these
solitonic states saturated the so-called Olive-Witten limit, and
became BPS states, inmune to quantum corrections; there were ideal
objects for studying the strong  coupling limit of several
(eventually, all) string theories. \\

The strong ($S$) coupling limit of the five superstring theories
could then be described: \\

i) The $IIB$ theory dualizes to itself: the role of the
dilaton/$RR$ axion scalars and the fundamental/$RR$ 2-forms are
crucial to show this; besides, the theory exhibits \textit{D-branes}
as the sources of the $RR$ even-dimensional $p$-forms: in fact, there
are branes in all odd dimensions up to $9$. The strong duality
group was conjectured to be the infinite discrete modular group
$SL(2, Z)$, which will play an important role in the first version
(Vafa \cite{Vaf96}) of  $F$-Theory (see below). \\

ii) The Type $I$ theory is $S$-dual to the Het-Ortho theory
(Witten-Polchinski \cite{WittP96}): recall both have the same
full spectrum. \\

iii) But the two big surprises were the relations of strings with $11D$
Sugra: The strong limit of the $IIA$ theory in the mass zero
regime gives exactly the particle content of $11D$ Sugra, as shown
by Witten (March 1995, \cite{W95}); this was really the crucial
starting paper in $M$-theory; but the name, ``$M$'', were $M$
stands for Membrane, Magic or Mystery according to taste, was not
endorsed until October, 1995 (\cite{Schw95}). In other words: the
$11D$ Sugra theory compactified in a large circle becomes the massless
(strong) limit of the $IIA$ theory, and the extant $11D$ membrane shrinks
to the $10D$ string by double dimensional reduction. \\

iv) $11D$ Sugra compactified in a segment $D^1 =S^1/Z_2$ is very likely equivalent to
the Het-Excep theory (Witten and  Ho\v{r}awa \cite{WitH96}); namely the
boundary is acceptable only supporting gauge groups, indeed one $E_8$
group at each corner of the segment; now as the segment shrinks,
and the $11$ dimensions become ten, we reproduce the Het-Excep
theory;  at least, as the authors emphasized, \textit{if} there is
a limit of the  $11D$ Sugra/segment construct, {\it then} it has to be the
$m=0$ sector of the H-E theory. In another development, Witten
(\cite{Witten00}) argued that if the segment is much larger than
the Calabi-Yau space (which links the H-E corner with the $4D$
world), the universe appears fivedimensional, and one can even
make the dimensionless gravitational constant $G_{N}  E^2$
coincide with the GUT value ($\approx 1/24$) of the other three.
\\

Branes can appear as solitons as well as sources for the $RR$
forms, in the same way that the fundamental (or $F$) string is the
source for the $NS$ 2-form. For example $11D$ Sugra supports the
membrane (Townsend) and \textit{also} a dual $5$-Brane
\cite{Guven92}: duality for branes occur at the field strength
level; e.g. the $11D$ membrane ($p=2$ extended object) couples to
the $3$-form $C$ of before, and it is its field strength, a $4$-form, which
dualizes to a $7$-form, supported by a $5$-brane. And in $IIB$
theory, the $RR$ radiation forms, the $0$-, $2$- and $4$-
(selfdual) potential forms require as $RR$ charges or sources
Polchinski\'{}s $D$-$p$-Branes, which are extended objects acting
as terminals of open strings. \\

There is a further equivalence, between $IIB$ and Type $I$; it is
achieved through the so-called $\Omega$ (parity) projection, 
forming unoriented strings.
For dualities, Branes and $M$-theory the reader should consult the
books \cite{Pol98} and \cite{John03}, and the magnificent
collection of original papers gathered in \cite{Duff99}. \\

The full set of relations between the different types of
Superstring Theory together with $11D$ Supergravity are
schematized in the famous hexagon \\

\begin{equation}\label{eq:20}
\begin{array}{clcr}

 &   11D \ Sugra &  \\
 &   &   \\
IIA &       &  H-E \\
    &       &          \\
IIB &        &   H-O  \\
    &       &    \\
    &   \ \ Type I   &
\end{array}
\end{equation}

\section {Some difficulties with $M$-theory.}

\subsection {Some developments in $M$-theory.}

 $M$-theory has not lived up
its expectations. The theory is so badly defined, that it is even
difficult to say it is wrong! In the elapsed ten years, it has
not advanced very much. At the begining of it, ca. 1996, many new
avenues were explored, by Witten and others, with no real
progress; we just quote some of them: \\

1) Matrix models \cite{Suss97} \\

2) AdS/CFT correspondence \cite{Malda98}  \\

3) Nonconmutative geometry \cite{SeiW02} \\

4) F-theory \cite{Vaf96}, which we shall develop below in Sect.
3. \\

5) K-Theory \cite{FreedW03}  \\

6) $E_{11}$ symmetry \cite{Wes01} \\

7) Topological strings and twistors \cite{Witt4}\\
 
In ``$M$-theory as a Matrix model'' \cite{Suss97} the authors
suggest a precise equivalence between $11D$  $M$-theory and the
color $N = \infty$ limit of a supersymmetric matrix mechanics
describing particles (as $D$-$(p=0)$-Branes). \\

The Anti de Sitter - Conformal Field Theory correspondence
(AdS/CFT; \cite{Malda98}, see also \cite{Malda00}), is an important
development, independently of its future use in a ``final''
theory. In the spirit of the Holographic Principle of  \'{}t Hooft
and Susskind, it is argued that the large $N$ ($N$ colors) limit
of certain $U(N)$ conformal field theories includes a sector
describing supergravity in the product of Anti-de Sitter spacetime
and a sphere: the light cone where the CFT sits can be seen as the
boundary of the AdS space. In particular, restriction of
strings/$M$-theory to Anti de Sitter spacetime is dual to
superconformal field theories. Roughly speaking, concentrated gravity in $D+1$
dimensions exhibit the degrees of freedom in the light cone with
$D$ dimensions, because the ``bulk'' lies inside the black hole
horizon, and therefore it is unobservable. \\

Nonconmutative geometry is a new advance in pure
mathematics due to A. Connes \cite{Conn93}, \cite{JMGB04}; it is
interesting that position or momentum operators conmutation rules became
nontrivial in presence of magnetic fields or the 2-forms $B$ of
string theory. This is the inicial thrust of the long
Seiberg-Witten paper, \cite{SeiW02}. \\

$K$-theory as applied to $M$-theory \cite{FreedW03} considers the
three-form $C$ as a kind of local ``connection'', and it migh be
that $C$ as a 3-form is not globally defined; the gauge group is
the group $E_8$, which seems to play an important role in
$M$-theory (but it is explicit only in the H-E corner). 
The right mathematical tool to deal with this
situation is the so called $K$-Theory, which classifies vector
bundles. See also \cite{DiacF04}.  \\

P. West has developed an interesting scheme \cite{Wes01}
based in the infinite Lie algebra called $E_{11}$
; it is a triple extension of $E_{8}$, beyond the affine and
hyperbolic Kac-Moody cases. For a recent treatment see \cite{Wes04}. \\

For a recent review of topological strings see \cite{Mar5}

\subsection {Stringy Microscopic Black Hole Entropy.}

Perhaps the only ``success'' of $M$-theory so far is the accurate
counting of microstates, and hence of the entropy, in some types
of Black Holes (BH) by Strominger and Vafa \cite{StroV96}. \\

In gravitation the equivalence principle forces the existence of
a horizon for the field of a point mass (the same phenomenon does
not occur for a point charge inspite the same $r^{-2}$ force law).
In quantum mechanics the BH radiates \cite{Hawk75}, and it has
therefore temperature and entropy: the later is the horizon
area/$4$, in Planck\'{}s units. The above authors were able to compute
this as related to the number of microstates, as one should do of course in
the statistical mechanics interpretation of thermodynamics. \\

Let us mention that recently Hawking \cite{Hawk05} has withdrawn
his previous claim that there is information loss in BH radiation:
there might be subtle correlations in the black hole temperature
radiation, ``remembering'' what came into the black hole in the
first place.  \\

 \subsection {Difficulties with $M$-theory.}

Inspite of these separate advances (and some others we omit) the
theory languished, both for lack of new stimulus, as for being
unable to complete its deficiencies. Among the unsatisfactory
features of $M$-theory as first established we can quote: \\

1) There is no clear origin for the $IIB$ theory: indeed, the
relation within the hexagon is by means of $T$-duality, which
means going to nine (or less) dimensions; one should hope to
obtain the proper $IIB$ theory in $10D$ in some limit of the (future)
$M$-theory. \\

2) Even the Heterotic Exceptional corner of the hexagon is a bit
far fetched if related to the $11D$ Sugra: where do the two $E_8$
groups come from? As we said, Witten and Ho\v{r}awa were careful
enough to state ``IF the 11D/Segment reproduces a string theory,
THEN it has to be the H-E corner''. Of course, the two gauge
groups $E_8$ at boundaries are forced by anomaly cancelation, but
one does not ``see'' them directly in the $M$-theory. \\

This argument is really more powerful: it is only the $IIA$ corner
which fits reasonably well with the $11D$ theory; indeed, neither
the $H$-$O$ theory nor the Type $I$ come directly from
$M$-theory, one has to recur to $T$-duality; 
also, even as regards the $IIA$ theory, see
point 5) below. \\

3) There is lack of a dynamical principle; in this sense string
theory, at least, is more conventional than $M$-theory:  excited
strings can be treated, at least perturbatively, as a Quantum
Theory; however, membranes and higher ($p > 2$) Branes have no
known quantization scheme. This is related to the dimensionless
character of the $2$-dim quantum fields, which allows very general
background couplings, and this is simply not true for membranes
etc. A related argument is this: for particles and strings, the
geodesic problem (minimal volume) in the ``Polyakov form'' is
equivalent to gravitation in one or two dimensions (Weyl
invariance is needed in the string case). However, this is no
longer true from membranes onward: although there is no really
graviton degrees of freedom in $3D$, gravitation is ``conic'', and
presents definite phenomena. \\

4) The maximal natural gauge group in $M$-theory, in its $11D$
version, seems to be the ${8 \choose 2}= 28$-dim orthogonal $O(8)$
group, generated by the 28 massless gauge fields down to $4D$.
But this group is too small, and unable to accomodate the minimal GUT group of the
standard model, $O(10)$ ($SU(5)$ is insufficient to account for
massive neutrinos); there are possible way outs, like composite
fields, $SU(8)$ as gauge group, etc., but none really very convincing \cite{Nico89}. \\

5) There is the so-called massive $IIA$ theory \cite{Romans86}
which again does not fit well with $M$-theory. It is a new version
of the $IIa$ nonchiral supergravity in ten dimensions, in which
the two-form $B$ ``eats'' the one-form $A$ (the vector field) and
grows massive: it is exactly the Higgs mechanism one degree
further; but then the analogy with the reduced sugra in $11D$ no
longer subsists, and hence it corresponds to no clear corner in
the $M$-theory hexagon. \\

\section {Forward with  $F$-theory.}

\subsection {The Original Argument.}

Cumrum Vafa seems to have been the first, back in  February 1996
\cite{Vaf96}, to advertise an ``$F$'' Theory, in $12$ dimensions
with $(-2, +10)$ signature, after the first introduction of the
$M$-theory by Witten in March, 1995 (although the name,
$M$-theory, was given a bit later), and as an extension of the
same; for another early hint on $12D$ space see \cite{Hull96}. \\

The argument of Vafa was related to the $IIB$ superstring theory;
as it lives in $10D$, it cannot come directly from $M$-theory in
$11$ dimensions: the two possible one-dimensional
compactifications from $11D$ were on a circle $S^1$, giving the
$IIA$ theory, and in a segment $D^1$, giving the Het-Excep theory;
besides, there were no questions of any strong coupling limit: as
we said, the $IIB$ string was a case in which selfduality under
$g_s \rightarrow 1/g_s$ was proposed, because the two scalars
(dilaton and axion) make up a complex field $z$, which transform
homographically under the discrete residue of $SL(2, R)$, a well
known invariance group (although noncompact!) of $IIb$ Sugra; the
two $B$ fields (fundamental and RR) also transform naturally under
this $SL(2, R)$ group; Townsend was the first to propose then that
a discrete $SL(2, Z)$ subgroup remained, and it was then obvious
that the duality included inversion of the string coupling, $g_s
\rightarrow 1/g_s$, where $g_s$ is the exponential of the vev of
the dilaton $\phi$.  \\

The point of Vafa was that the group $SL(2, Z)$, the so-called
modular group, was the moduli group for a torus: the inequivalent
conformal structures in the torus are labeled by the modular
group. And Vafa, of course, interpreted this torus, with metric
$(-1, +1)$, as a compactifying space from a $(-2, 10)$ signature
space in twelve dimensions; the name ``$F$-theory'' was proposed
by Vafa himself, meaning probably ``father'' or ``fundamental'';
the argument for increasing one time direction is subtle and we
shall show it more clearly later below. So the
idea is that the $IIB$ theory on space ${\cal{M}}_{10}$
comes from an elliptic fibration with
fiber a $2$-torus, in a certain $12D$ space: symbolically

\begin{equation}\label{eq:2}
T^2  \  \rightarrow   \    {\cal{V}}_{12} \ \rightarrow \    {\cal{M}}_{10}
\end{equation}

Then, $F$-theory on $ {\cal{V}}_{12} $ is equivalent to $IIB$ on
 $ {\cal{M}}_{10} $.

\subsection { Dynamical Arguments for $F$-theory.}

There are several other arguments, mainly aesthetic or of
completion, in favour of this $F$-theory. We express all of them
rather succintly, as none are really thoroughly convincing. However,
there are so many that we believe taken together they give some force
to the idea of a 12-dimensional space with two times; 
for the particle content in this space see a proposal later. \\

1) $IIB$ theory really comes from $12$ dimensions, with toroidal
compactification; this was the original argument of C. Vafa
\cite{Vaf96}. The space ${\cal{V}}_{12}$ admits an elliptic
fibration, and the quotient is the frame for the $IIB$ theory. Most of 
compactifications from strings ($10D$) or $M$-theory (in $11D$)
can be carried out from $12D \ F$-theory \cite{Vaf96}, \cite{Sen96}. \\

2) There is a Chern-Simons (CS) term in the $11d$ Sugra
lagrangian, together with the conventional kinetic terms for the
graviton, gravitino and $3$-form $C$, and another ``Pauli type''
coupling, see e.g. \cite{Crems} \\

\begin{equation}\label{eq:29}
    {\cal L} = ... \ \  + C \wedge dC \wedge dC + ...
\end{equation} \\

Now a CS term can be understood as a boundary term, hence claiming
for an extra dimension interpretation $(dC)^{3}$ \cite{EAlva00}. This favours
the interpretation of the group $E_8$ as a gauge group in
$M$-theory \cite{DiacF04}. \\

3) There is a $(2, 2)$ Brane extant in the Brane Scan, once one
allows for some relax in supersymmetry dimension counting. In fact, in
the brane scan of Townsend, if one insists on $(1, D-1)$ signature,
one finds the four series of extended objects (and their duals),
ending up with the $11D \ \ p=2$ membrane \cite{Freed04}. By relaxing the signature,
but still insisting in supersymmetry (in the sense of bose-fermi
matching), one encounters a few new corners \cite{BainDf88}; Susy
algebras in this most general context were already considered in
\cite{Vanvan82}. Indeed, there is a $(2, 2)$ membrane living in
$(2, 10)$ space. The membrane itself was studied carefully in
\cite{Perry97}; by doubly dimension reduction, this $(2,2)$
membrane supposes to give rise to the string in $IIB$-theory. \\

4) The algebra of $32$ supercharges of $11D$ Sugra still
operates in $12 = (2,10)$ dimensions, with the (anti-)conmutation relations

\begin{equation}\label{eq:3}
\{Q, Q \} = 2-form + 6^{\pm} -form
\end{equation}

\noindent so dim $Q = 2^{12/2 \ -1}$ and  $528={12 \choose 2} + {12 \choose 6}/2$
(for the interpretation see later). \\

Now, in $11D$, dim $Q = 2^{(11-1)/2} =32$ (type $+1$, real),
and the superalgebra is 

\begin{equation}\label{eq:4}
\{Q, Q \} = P_{\mu} + Z^{(2} + Z^{(5}  \ \ \ \ \ \ 32 \cdot 33 /2 = 11 +
55 + 462
\end{equation} \\

\noindent namely, $11$-dim translations plus a two-form and a
$5$-form, understood as central charges. But the $11D$ 
superalgebra clearly comes from the simpler $12$-dim algebra of above: 
the $2$-form gives $1$-form and $2$-form, 
and the selfdual $6$-form gives rise to the five-form.
The signature must be $(2, 10)$, which gives $0$ mod $8$ for the
type of $Q$; in case of $(1, 11)$ signature is $2$ mod $8$: the charges would be complex, or
$2 \ \cdot \ 32= 64$ real: twelve dimensional two-times space is 
maximal for 32 real supercharges. \\

On the other hand, the direct reduction from $12D$ 
space to $10D$ via a $(1, 1)$ torus would yield the $IIB$ 
string from the $(2,2)$ membrane (as noted above)
and the self-four form $D^{\pm }$ from the seldfual $6$-form. \\

5) We have remarkable relations in dimensions $8D, \ 9D, \ 10D$: effective
dimensions of a $(1, 9)$ string theory, $M$-theory in $(1, 10)$
and $F$-theory in $(2, 10)$ as regards supersymmetry.
The fundamental supersymmetry extant in $8$ effective
dimensions in string theory is $8_{v} - 8_{s}$
between the vector ($v$) and the spinor ($s$), and squaring

\begin{equation}\label{eq:30}
(8_{v} - 8_{s}) \cdot (8_{v} - 8_{s'}) = h \  - \Psi \  + C = 44 - 128 + 84
\end{equation} \

That is, the square fits in the content of $11d$ Sugra with $9$
effective dimensions. \\

But another square  {\# }$( h - \Psi + C)^{2} = 2^{15}-2^{15} $ gives a $27$-plet 
in $D_{eff} =  10$  (or $(2, 10)=12D$) as we shall see later, because this is
related to our proprosal for the particle content of $F$-theory. The fact that the
square of the fundamental irreps of Susy in $8$ effective dimensions fits nicely
in irreps of $9D$, and the square again fits in irreps of a effective $10D$
theory, is most notorious and unique, it is certaintly related to
 octonion algebra, and it was first noticed by I. Bars in \cite{Bars88} before
the $M$-theory revolution. \\

6) The content of $11D$ Sugra is related to the symmetric space
(Moufang projective plane over the octonions)

\begin{equation}\label{eq:5}
OP^2 = F_4/Spin(9)  \ \  
\end{equation}

\noindent where $O(9)$ is the massless little group in $11D$; 
this was shown by Kostant \cite{Kost00}. There is a
natural extension by complexification \cite{Atiy01} to
 the space

\begin{equation}\label{eq:6}
OP_{C}^{2} = E_6/Spin^{c}(10)
\end{equation}

\noindent related naturally to the
$12 = (2, 10)$ space as $Spin^{c} = Spin(10) \times_{/2} U(1)$, where
$O(10) \times O(2)$ is
the maximal compact group of the $O(2, 10)$ tangent space group in $12D$. \\

We shall explain this in detail in section four. It will be enough
to remark here that now the candidate GUT group is $O(10)$, and there is no
problem with fitting the SM group ($U( \ 3, \ 2, \ 1)$) within it. \\

7) Compactification from $11$ dimensions to $4$ is preferable
through a manifold of $G_2$ holonomy in order to preserve
just $N=1$ Susy in $4D$; $G_2$ is a case of $7D$
{\it exceptional} holonomy, the only other being $Spin(7)$, acting in $8$
dimensions, very suitable for our $12 \  \rightarrow 4$ descent
 (this was already noticed by \cite{Vaf96}); again, the $11D$ case generalizes
naturally and uniquely to $12$; and we have the nice split $12=4 + 2 \cdot 4$.
Indeed, it seems that an argument like 4) can be also made here.
Trouble is, we really need  $8$-dim manifolds with $(1, 7)$
signature and exceptional holonomy, which are not yet fully studied. \\

\subsection {Numerical Arguments.}

These are really ``numerological'', and we include them for lack
of a better reasoning, and also for some kind of ``completeness''. \\

1) $O(2, 10)$, the tangent space group, is the largest group of
the Cartan coincidences; namely, it is

\begin{equation}\label{eq:7}
O(2, 10) = Sp(2, Oct)
\end{equation}

\noindent in a definite sense that we can briefly comment:
namely the groups $Sp(1,\ 2;\ K)$ where $K$ are the reals, complex, quaternions
or octonions, are equivalent to some Spin groups. This is part
of the so called Cartan identities \cite{Sud}; the simplest of these is
 $Sp(1, R) = Spin(1, 2)$, and the largest is
 $Sp(2, Oct) = Spin(2, 10)$, pertinent to $F$-theory. \\

2) In the $12 = 4 + 8$ split, all spaces are even, and
even dimensional spaces have integer
dimensions for bosons and halfinteger for fermions, which seems to
be the natural thing; for odd dim spaces, it is the other way
around. \\

3) $78 =$ dim  $E_6 =$ dim ( Poincar\'{e} or (A)dS  $(2, 10)) =$ dim   $O(13) =$ dim 
$Sp(6)$ is the only dimension, after the $A_1 = B_1 = C_1$ identities,
with triple coincidences of group dimensions; does this hint towards a relation
between $E_6$, as a GUT group, with the $(2,10)$ space of $F$-theory? \\

\subsection {Unusual characteristics of $F$-Theory.}

Actually, $F$-theory as presented here is very different from the
usual theories; in particular, as the $\{Q,Q \}$ (anti)conmutator does
not contain the translations, there is no ``Poincar\'{e}'' algebra,
and consequently there is no direct concept of mass as eigenstate
of momentum, and in particular no massless limit.\\

 How does one understand this? In any arbitrary (pseudo-)riemannian manifold
the only general structural feature is the structure group of the
tangent bundle, or in physical terms the tangent space group,
which is the (pseudo-)orthogonal group; in our case of the $(2,
10)$ space, it is $O(2, 10)$; so it is gratifying that indeed, 
the supercharge algebra (14) gives just this generator, as a two-form.
The concrete shape of the space is left unanswered, for the moment. \\

As for the selfdual $6$-form in (14), it is hoped it will be related 
to the matter content, in the same sense as the central charges
in $11D$ Supergravity are related to membranes.
Is it possible to relate this selfdual $6$-form to the extant ($2, 2$)-Brane? \\

How do we incorporate two times in a theory of physics, in which
the arrow of time is so characteristic? At face value, there are
two ways out: either one of the times compactifies, so possible
violations of causality are of the order of Planck\'{}s length, or
there is a gauge freedom to dispose of one of the times; I. Bars
\cite{Bar00} favours the second, but there are also consistent schemes
with the first alternative. \\

\section {Euler  Multiplets.}

\subsection {Euler Triplet as the ${\bf 11d}$ Sugra particle content.}

In 1999 P. Ramond realized \cite{Ram99} that many irreducible
representations (irreps) of the $O(9)$ group, including the 
Sugra irrep

\begin{equation}\label{eq:8}
h - \Psi + C = 44  -  128  +  84
\end{equation} \

\noindent i.e. graviton, gravitino, $3$-form, can be grouped in
triplets with many coincidences, besides the ``susy'' aspect (all
but one of the Casimirs correspond, etc.). The underlying mathematics was
cleared up by B. Kostant \cite{Kost00}, and goes with the name of
``Euler triplets'' (later multiplets). It is related to the
symmetric space

\begin{equation}\label{eq:9}
OP^2 = F_4/O(9)
\end{equation}

\noindent which defines the octonionic (or Moufang) plane. Namely
the spin irrep of the orthogonal group acting in $R^n$
with $n=$ dim $F_{4}$ - dim $O(9) \ =16$, splits in $3$ irreps 
of $O(9)$ because the Euler number of projective planes is three: 
$\chi (OP^2)=3$; it is supersymmetric 
because the Dirac operator has index zero, and this measures 
the mismatch between left- and right- nullspinors. \\

We shall not delve in the proofs, only on the construction. Let
$G\supset H$ be a pair: Lie group and subgroup, of the same rank,
with $G$ semisimple and $H$ reductive (i.e. $H$ can have abelian factors). The
$Pin(n) = \Delta_{L} + \Delta_{R} $ group in the difference of dimensions 
 $n = dim \  G - dim \  H$ suggests a Dirac operator $D$: left spinors to
right spinors (these manifolds are even-dimensional,
have $w_2=0$ or are $Spin^{c}$, so they admit spinor fields). Then
ker $D$ and coker $D$ split each in representations of $H$, with 
a total of $r = \chi (G/H)$ irreps of $H$. Notice dim $Pin(n)=2^{n/2}$ and
dim $\Delta_{R, L}  = 2^{n/2 -1}$. Now this Euler number can be computed from
the Weyl group of the $(G, H)$ pair:

\begin{equation}\label{eq:Z}
\chi (G/H) = \frac{\# \ Weyl \ group \ of \ G} {\# \ Id. \ of \ H}
\end{equation}

So there is a kind of supersymmetry associated to some
homogeneous spaces. The matching is as follows:

\begin{equation}\label{eq:Y}
\Delta_{L} - \Delta_{R} =  \sum_{i=1}^{\chi} \ \pm D_{i}(H)
\end{equation}

\begin{equation}\label{eq:X}
2^{n/2 -1} - 2^{n/2 -1} = \sum_{i=1}^{\chi} \ \pm \ dim \ D_{i}  (H)
\end{equation} 

In fact, each representation of the higher group $G$ generates
 a similar multiplet of $\chi$ irreps of $H$;
 the above construction corresponds to the Id irrep of $G$; see \cite{Kost00}. \\

We shall use a complexification of the Moufang plane 
as a proposal space for the particle content in $F$-theory; 
before this we give simple examples of the Ramond-Kostant construction.\\

\subsection  {Some examples of Euler multiplets.}

There are many example of this Ramond-Kostant construction; we
shall exemplify only two of them. Consider the complex projective
space

\begin{equation}\label{eq:10}
CP^n = SU(n+1)/U(n);
\end{equation}

\noindent  the homology of $CP^n$ is $(1, 0, 1, 0, 1,
...)$, so $\chi(CP^n)=n+1$; on the other hand, dim $\Delta_{L, R}(2n)=
2^{n-1}$. The ``supersymmetric partition'' is e.g. for $n=4$

\begin{equation}\label{eq:11}
\Delta_L - \Delta_R = [1^{0}]^{+2} - [1]^{+1} + [1^2]^0 - [1^3]^{-1} +
[1^4]^{-2}
\end{equation}

\noindent (or $8 - 8 =1 -4 + 6 - 4 + 1$), with the usual convention: $[D]^q$ is the irrep of
$U(n)$ in the form $[D]$, an irrep of $SU(n)$, and $q$, a label for
the $U(1)$ factor, where $U(n)=SU(n) \times_{/n} U(1)$. In eq. (26)  only the
fully antisymmetric irreps $[1^r]$ enter, and $[1]$, dim $=4$, is the fundamental
irrep of $SU(n=4)$. Notice for $n$ odd 
 ``Supersymmetry'' is just Poincar\'{e} duality. \\

    Another more sophisticated example is this \cite{RamB}. Take $G=SO(10)$ 
and $H = SO(8) \times SO(2)$. Then dim $G/H=16$ and Euler $\# = 10$. The
dimensions of the split multiplet is

\begin{equation}\label{eq:18}
\\ 1^{2} - 8^{3/2} + 28^{1} - 56^{1/2} + (35^{+0} + 35^{-0}) - 56^{-1/2}
+ 28^{-1} + 8^{-3/2} + 1^{-2}
\end{equation} \\

\subsection {The proposal.}

The original discovery of Ramond was related to the Sugra $11D$ multiplet,
associated to the Moufang plane (this association was first seen
by Kostant). \\

Now Atiyah and Berndt \cite {Atiy01} have shown that the first row
of groups in the so-called magic square

\begin{equation}\label{eq:12}
O(3) \ \  U(3) \ \  Sp(3) \ \  F_4
\end{equation}

\noindent associated to projective planes (e.g. $RP^2 = O(3)/O(1)
\times O(2)$, etc.) can be complexified in a precise mathematical
form; thus the second row is $U(3), U(3)^2, U(6)$ and $E_6$; the
complexified symmetric spaces are also well studied; in particular
the fourth is

\begin{equation}\label{eq:13}
OP^2 \rightarrow OP_C^2, \  \  F_4/SO(9) \rightarrow E_6/SO(10)
\times SO(2)
\end{equation}

It is natural to extend the Sugra multiplet, associated to $O(9)$
as the massless little group of the Poincar\'{e} group in $(1, 10)$,
to the multiplet associated to the complexification of the
Moufang plane, namely

\begin{equation}\label{eq:14}
 Y: = \frac{ E_6}{SO(10) \times SO(2)}
\end{equation}

This is a remarkable space: it is symmetric, hermitian, and exceptional; 
the subgroup should really be $Spin(10) \times_{/2} U(1) = Spin^{c}(10)$ \\

 How do we connect this with $F$-theory? Because $O(10)
\times O(2)$ is the maximal compact subgroup of the tangent group
$O(2, 10)$ in our $12D \ F$-theory, as we said above! In ordinary, 
($1$, $D-1$) spaces, the massive group is $O(D-1)$ and the massless $O(D-2)$.
But here, with a ($2, D-2$) space, we suggest naturally to take the maximal compact 
subgroup, namely $O(2) \times O(D-2)$. \\

The corresponding multiplet of particles for $Y$ is huge: 
the Euler number is $27$, 

\begin{equation}\label{eq:15}
\chi(OP_C^2) = \# Weyl(E_6)/ \#Weyl(O(10)) = 51840/1920 = 27
\end{equation} \

\noindent and the split $\Delta_{L}(32) - \Delta_{R}(32) $
 is $2^{15} - 2^{15}$ or $32768 - 32768$. \\

The particle content has been calculated by I. Bars \cite{Bars88}.
 Because the complex nature of $Spin(10)$, dim $\Delta(10) = 16$,
 it helps to consider the chain

\begin{equation}\label{eq:15}
Spin(8) \subset Spin(9) \subset Spin(10) \subset SU(16) \subset
SO(32)
\end{equation}

\noindent and it is given in the following Tables. (We omit the $q$ label
associated to the abelian part SO(2), and write only half of the Table, 9
out of the 17 entries, to be completed by duality, 
as $ n \choose k$ = $n \choose n-k $). \\

\begin{table}
	\begin{center}
		\begin{tabular}{ | l | r | r | r | r | r | r | r | r | }
			\hline
			$O(32)$   &               &          &  $\Delta_{L}$ &   $2^{15}$         &                   \\ \hline
			$SU(16)$  &  Id=[$1^{0}$] & [$1^2$]  & [$1^4$]       & [$1^6$]     & [$1^8$]            \\  \hline
			SO(10)    & scalar        & 3-form   & Weyl + 1050   &   4312+3693 & 4125+8085+660              \\ \hline
    dim      &   1           &   120    &  1820         &  8008 & 12870   \\  \hline
  \end{tabular}
			\caption{Putative Bose Particle content of $F$-Theory}
	\end{center}
\end{table}

\begin{table}
	\begin{center}
		\begin{tabular}{ | l | r | r | r | r | r | r | r | r | }
			\hline
			$O(32)$   &               &           &  $\Delta_{R}$ &   $2^{15}$      \\ \hline
			$SU(16)$  &  16=[$1$]     & [$1^3$]   & [$1^5$]       & [$1^7]$          \\  \hline
			SO(10)    & spinor        & h-g-ino   & 672+3696      &   8800+2640    \\ \hline
    dim      &   16          &   560     & 4368    & 11440  \\  \hline
  \end{tabular}
			\caption{Putative Fermi Particle content of $F$-Theory}
	\end{center}
\end{table}         

This is a huge multiplet; notice it contains an scalar, a spinor, a 3-form, 
 the Weyl tensor $[2^{2}]_{tl}$, and the hypergravitino (h-g-ino),
 $\Psi_{\mu \nu \alpha } $,
 plus more exotic representations. It does not contain 
gravitons nor gravitinos. There is no clear-cut between massive
 and massless particles. 
Evidently, there should be a way to cut down the number of
states in order to obtain some sensible physics; 
we have no idea how this can be done, out of the simple remark that in the
standard minimal
supersymmetric model (with no gravitons) the number of states is $128 - 128$, 
the square root of those obtained here. Perhaps a square 
root {\it different} of that obtaining the $11D$ content
 from the $12D$ is singled out.
 
\section {Outlook.}

What will be the physics of this putative $F$-theory? If true,
it will be a theory very different from the usual ones. We focus
here in one particular aspect, the projective spaces $OP^{2}$ and
$OP_{C}^{2}$; notice the particle content of the second does contain
neither the graviton nor the gravitino; could this be an indication
that projective geometry will take the place of the riemannian (metric)?
After all, the second is a particular case of the first; also, much of the dynamics
in $M$-theory, which we hope will persist in $F$-theory, has to do with
Branes: intersection, pile-up, $D$-type, etc. This reminds one of the 
operations in projective geometry, with inclusion and intersection of
subspaces; we take as a hint the fact that both the Sugra triplet
and the $27$-plet considered here single out projective spaces.\\

	Even the usual lagrangian formalism used so far becomes suspect: 
dualities in $M$-theory make two different lagrangians to produce the same physics,
 which is indicative
of a more general formalism, one of the forms of which is lagrangian: it is like
the theory of transformations in quantum mechanics, which is the
invariant formalism, whereas the coordinate or momentum representations are just
partial aspects.\\

	We finish by making the obvious comparison with the times of the Old
Quantum Theory, 1913-25: there were some results, some recipes, but absence of a
full dynamical theory, which came only with the matrix mechanics of Heisenberg in 1925. 
Perhaps something of this kind of surprise is in store for us. 
It is in this spirit that we offer these somewhat wild speculations
on the space manifold and particle content of the future theory. \\\

{\bf Acknowledgements}. The essential content of this paper was presented 
at Seminars at MPI (Munich), Theory Group (Austin, TX), U.C. Madrid,
 and other places; the author is grateful for discussions with P. Ramond,
C. Gomez, P. Townsend, G. Gibbons, P. West, J. Distler and V. Kaplunowski. 
I would like also to thank the MCYT (Spain) for grant FPA2003-02948.

\vfill \eject

\end{document}